\documentclass[10pt]{iopart}

\expandafter\let\csname equation*\endcsname\relax
\expandafter\let\csname endequation*\endcsname\relax
\newcommand{\newblock}{}  

\usepackage{amsmath}   

\usepackage{lmodern}    
\usepackage{bm}
\usepackage{microtype}

\usepackage{soul}
\usepackage{color}   

\usepackage{graphicx}
\usepackage{booktabs}       
\usepackage{tabularx}
\usepackage[hyphens]{url} 

\usepackage{SIunits}
\usepackage{sistyle}
\newcommand{\si}[1]{\SI{}{#1}}

\usepackage[square,numbers,sort&compress]{natbib}

\usepackage[%
  urlcolor=darkblue,%
  citecolor=darkblue,%
  linktoc=page,%
  breaklinks,%
  hyperfootnotes=false,%
  hidelinks,
  ]{hyperref}

\usepackage{cleveref}
\crefname{equation}{}{}
\Crefname{equation}{Equation}{Equations}
\crefname{figure}{figure}{figures}
\Crefname{figure}{Figure}{Figures}
\crefname{table}{table}{tables}
\Crefname{table}{Table}{Tables}
\crefname{appendix}{}{}
\Crefname{appendix}{}{}

\renewcommand{\vec}[1]{\bm{#1}}
\let\oldhat\hat
\renewcommand{\hat}[1]{\oldhat{\bm{#1}}}

\makeatletter

\begingroup
\catcode`_=\active
\catcode`^=\active
\protected\gdef_{\@ifnextchar'\subtextrm{\@ifnextchar"\subtextsc\sb}}
\protected\gdef^{\@ifnextchar'\suptextrm{\@ifnextchar"\suptextsc\sp}}
\endgroup
\def\subtextrm'#1'{\sb{\textrm{#1}}}
\def\suptextrm'#1'{\sp{\textrm{#1}}}
\def\subtextsc"#1"{\sb{\textsc{#1}}}
\def\suptextsc"#1"{\sp{\textsc{#1}}}
\AtBeginDocument{\catcode`_=12 \mathcode`_="8000}
\AtBeginDocument{\catcode`^=12 \mathcode`^="8000}

\renewcommand{\d}[1]{\text{d}{#1}}

\makeatother

\renewcommand{\hl}[1]{#1}             

\begin{document}



\title{%
  Conductivity and capacitance of streamers
  in avalanche model for streamer propagation
  in dielectric liquids%
  }


\author{
  I~Madshaven$^1$,        
  OL~Hestad$^2$,\\        
  M~Unge$^3$,             
  O~Hjortstam$^3$,
  PO~{\AA}strand$^1$
    \footnote{Corresponding author: \texttt{per-olof.aastrand@ntnu.no}}
  }

\address{%
  $^1$
  Department of Chemistry,
  NTNU -- Norwegian University of Science and Technology,
  7491 Trondheim, Norway
}
\address{%
  $^2$
  SINTEF Energy Research,
  7465 Trondheim, Norway
}
\address{%
  $^3$
  ABB Corporate Research,
  72178 V{\"a}ster{\aa}s, Sweden
}

\begin{abstract}%
%
Propagation of positive streamers
in dielectric liquids,
modeled by
the electron avalanche mechanism,
is simulated
in a needle--plane gap.
The streamer is modeled as an RC-circuit
where the channel is a resistor
and the extremities of the streamer
have a capacitance
towards the plane.
The addition of the RC-model
introduces
a time constant to the propagation model.
Increase in capacitance as a streamer branch propagates
reduces its potential,
while
conduction through the streamer channel
increases its potential,
as a function of the time constant of the RC-system.
Streamer branching also increases the capacitance
and decreases the potential of the branches.
If the electric field within the streamer channel exceeds a threshold,
a breakdown occurs in the channel,
and the potential of the streamer is equalized with the needle electrode.
This is interpreted as a re-illumination.
According to this model,
a low conductive streamer branch can propagate some distance
before its potential is reduced to below the propagation threshold,
and then the RC time constant controls the streamer propagation speed.
Channel breakdowns, or re-illuminations,
are less frequent when the channels
are conductive
and more frequent for more branched streamers.
%
\end{abstract}

\vspace{2pc}

\vspace{2pc}

%
%
\ioptwocol
%



\section{Introduction}\label{sec:introduction}{

When dielectric liquids are exposed to
a sufficiently strong electric field,
partial discharges occur
and a gaseous
channel
called a streamer is formed.
The many characteristics of streamers,
such as
shape,
propagation speed,
inception voltage,
breakdown voltage,
current,
and
charge
are described by numerous experiments performed
throughout the last half century
for various liquids and different experimental setups~%
{\citep{Devins1981,Torshin1995,Lundgaard1998cn8k5w,Kolb2008,Joshi2013cxms,Lesaint2016cxmf}}.
A streamer bridging the gap between two electrodes
can cause an electric discharge,
and a better understanding of
the mechanisms governing the inception and the propagation of streamers
is essential for the production of
e.g.~better power transformers
and the prevention of failure in such equipment
\citep{Wedin2014}.

Simulating a low temperature plasma in contact with a liquid
is a challenge in itself \citep{Bruggeman2016cxmn}.
For a propagating streamer,
phase change and moving boundaries
complicates the problem further
and simplifications are therefore required.
The finite element method has been used
in models simulating streamer breakdown
through charge generation and charge transport~%
\cite{Qian2005dmhk4f,Jadidian2014f55gj5},
even incorporating phase change~%
\cite{Naidis2016}.
However, the first simulations of streamer breakdown in liquids
applied Monte Carlo methods on a lattice~%
\cite{Niemeyer1984d35qr4},
and have since been expanded,
for instance by including conductivity~%
\cite{Kupershtokh2006d3d7d3}.
Another model use the electric network model
to calculate the electric field in front of the streamer,
which is used to evaluate the possibility for streamer growth or branching~%
\cite{Fofana1998bm4sd5}.

For positive streamers in non-polar liquids,
it is common to define four propagation modes
based on their propagation speed,
ranging from around
$\SI{0.1}{\kilo \metre \per \second}$
for the 1st mode
and exceeding
$\SI{100}{\kilo \metre \per \second}$
for the 4th mode.
2nd mode streamers
propagate at speeds of some
$\si{\kilo \metre \per \second}$
creating a branching filamentary structure
that can lead to a breakdown if the applied voltage is sufficiently high~%
\citep{Lesaint1998dcvzh7}.

Our previous work describes a model for propagation
of 2nd mode positive streamers
in dielectric liquids
governed by electron avalanches
\citep{Hestad2014,Madshaven2018cxjf}.
According to the model,
electron avalanches can be important for streamer propagation,
but
the results also showed
a relatively low propagation speed
and
a low degree of branching.
The streamer channel was represented by a fixed
electric field within the channel
between the needle electrode and the extremities of the streamer.
The model focuses on the phenomena occurring
in the high electric field in front of a streamer,
assuming these are the main contributors to the propagation.
However,
processes in the channel may be important
for the electric field at the streamer extremities,
which is why it is addressed in this study.
Here,
the channel is included
by considering its conductivity
as well as
capacitance between the streamer and the plane.

}


\section{Simulation model and theory}{\label{sec:model}

\begin{figure}[b]
    \centering
    \includegraphics[width=0.45\linewidth]{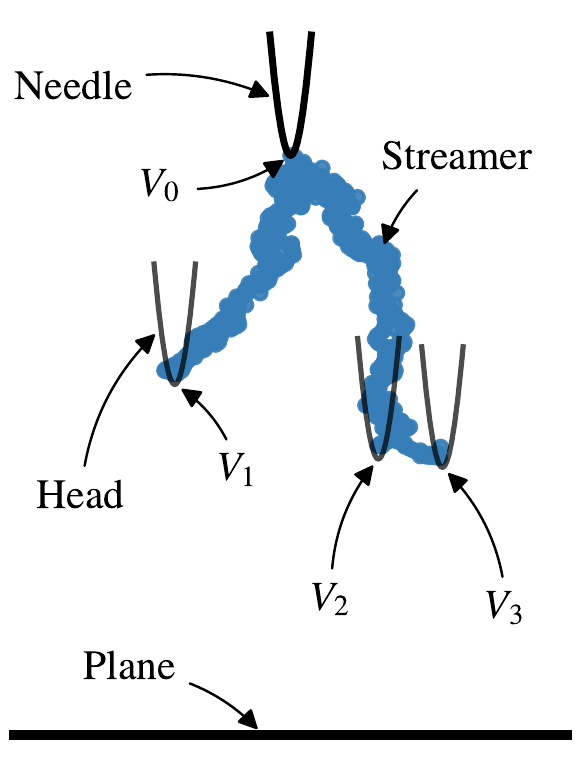}
    \hspace{1 ex}
    \includegraphics[width=0.49\linewidth]{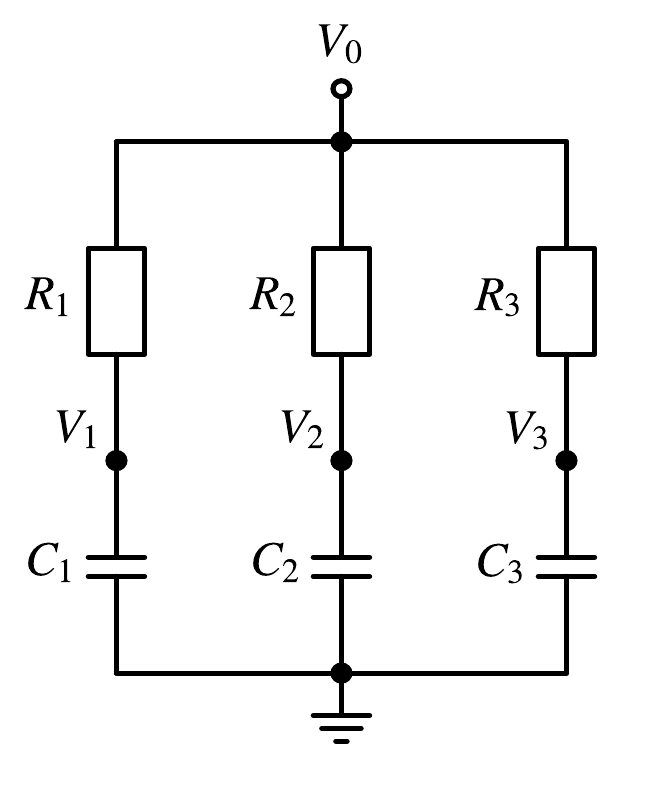}
    \caption{%
        (left) Illustration of the needle electrode,
        a branched streamer,
        and three streamer heads
        above a planar electrode,
        and
        (right) the equivalent RC-circuit.
        }
    \label{fig:streamer_rc}
\end{figure}

\subsection{Electron avalanche model}

We simulate streamer propagation in a
liquid-filled
needle--plane gap.
The needle is represented by a hyperboloid
and
the streamer is represented by a number of hyperboloidal streamer heads,
see \mbox{\cref{fig:streamer_rc}}.
Each hyperboloid $i$ has
a potential $V_i$ and an electric field $\vec{E}_i$.
\hl{A potential $V_0$ is applied to the needle when the simulation begins.
Since we here are interested in propagation rather than initiation of streamers,
a square wave with infinite risetime is applied.
The potential of each streamer head $V_i$
is dependent on the potential and capacitance
of the streamer (see \mbox{\cref{sec:el_pot_new_head}}),
and changes with time (see \mbox{\cref{sec:alg_streamer}}).
The method of calculation
gives a drop in potential between the needle tip and the streamer tip,
which is an important feature of the model.
The Laplacian electric field $\vec{E}_i$
is dependent on the potential $V_i$
and calculated using the hyperbole approximation%
~\mbox{\cite{Madshaven2018cxjf}}.}
The potential and electric field at a given position $\vec{r}$
is given by the superposition principle,
\begin{equation}
    V(\vec{r}) = \sum\limits_i k_i V_i(\vec{r}) \,
    \quad \text{and} \quad
    \vec{E}(\vec{r}) = \sum\limits_i k_i \vec{E}_i(\vec{r}) \,,
    \label{eq:Vr_Er}
\end{equation}
where \hl{the electrostatic shielding coefficients $k_i$ are} optimized such that
$V(\vec{r}_i) = V_i(\vec{r}_i)$,
\hl{i.e. the superposition of potentials
gives the correct potential at the tip of each head.}
Each head with $k_i$
lower than $k_'c'$ (shielding threshold) is removed
and heads closer than
{$d_'m'$} (head merge threshold) are merged%
~{\cite{Madshaven2018cxjf}}.
A number of anions,
given by the anion number density $n_'ion'$,
is placed at random positions in the liquid volume surrounding the streamer.
Anions are considered as sources of seed electrons,
which can turn into electron avalanches
if the electric field is sufficiently high.
The number of electrons $N_'e' = \exp (Q_'e')$ in an avalanche
increases each simulation time step $\Delta{}t$.
\hl{The change in $Q_'e' = \ln N_'e'$, $\Delta Q_'e'$, is given by}
\begin{equation}
    \Delta{}Q_'e' =
        E \, \mu_'e' \,
        \alpha_'m' \exp \! \left( - \frac{E_\alpha}{E} \right) \Delta{}t
    \,,
\end{equation}
where
$\mu_'e'$ is the electron mobility,
and
$\alpha_'m'$ and $E_\alpha$ are experimentally estimated parameters.
An avalanche is considered ``critical''
if $Q_'e'$ exceeds a threshold $Q_'c'$
(Townsend--meek criterion, \hl{$N_'e' = \exp Q_'e' > \exp Q_'c'$}).
Critical avalanches are removed,
replaced by a new streamer head.
The tip of the new streamer head is
positioned where the avalanche became critical,
and this way,
the streamer grows~{\cite{Madshaven2018cxjf}}.

The potential of the new head was set assuming
a fixed electric field $E_'s'$
in the streamer channel%
~\cite{Madshaven2018cxjf},
but here
the model is extended so that
the potential is instead calculated
by considering an RC-circuit.

\subsection{RC-circuit analogy for streamers}

A simple RC-circuit is composed of a resistor and a capacitor
connected in series.
When voltage is applied,
the capacitor is charged and its potential increases
as a function of time.
The time constant $\tau$
of an RC-circuit is
\begin{equation}
    \tau = RC \,,
    \label{eq:tau}
\end{equation}
where
$R$ is the resistance
and
$C$ is the capacitance.
Similarly,
the streamer channel
is a conductor with an associated resistance,
and
the gap between the streamer and the opposing electrode
is associated with a capacitance,
see \cref{fig:streamer_rc}.
This is a reasonable assumption when modeling a dielectric liquid
where the dielectric relaxation time
is long compared to the duration of a streamer breakdown%
~{\cite{Lesaint2016cxmf}}.


For a given streamer
length $\ell$,
cross-section $A$,
and
conductance $\sigma$,
the resistance is given by
\begin{equation}
    R = \frac{\ell}{A \sigma} \,.
    \label{eq:R}
\end{equation}
The resistance is proportional to the streamer length,
\hl{calculated as the straight distance from the needle to the streamer head.}
Also $A$ and $\sigma$ may change during propagation.
For instance,
during a re-illumination,
one or more of the streamer channels emit light%
~{\cite{Linhjell1994chdqcz}}.
This is likely
the result of the buildup of a strong electric field within the channel,
causing a a gas discharge within the channel,
increasing $\sigma$ and lowering $R$ significantly%
~{\cite{Dung2012czgj}}.
It seems reasonable to assume that the resistance
is reduced for some time after a re-illumination,
however,
measurements shows just a brief spike in the current,
typically lasting about $\SI{10}{\nano \second}$%
~\cite{Linhjell1994chdqcz},
which is consistent with the time scale for
charge relaxation of ions in the channel~\cite{Saker1996}.


The total charge of a streamer
can be found by integrating the current
and is in the range of $\si{\nano \coulomb}$ to $\si{\micro \coulomb}$
\cite{Massala1998,Ingebrigtsen2009fptpt5,Lesaint2016cxmf}.
The ``capacitance'' of the streamer can be approximated
by considering the streamer to be
a conducting half-sphere (slow and fine-branched modes)
or
a conducting cylinder (fast and single-branched modes),
which also enables the
calculation of the field in front of the streamer~%
\cite{Lundgaard1998cn8k5w,Massala1998,Top2002}.
We associate each streamer head with
the capacitance between itself and the planar electrode,
as illustrated in \cref{fig:streamer_rc}.
The capacitance then depends on the geometry of the gap between them,
and an increase in streamer heads increases the total capacitance of the streamer.
The capacitance for a hyperbole is applied for the avalanche model,
while
models for a sphere over a plane
and
a parallel plate capacitor
are included
here
as limiting cases.

The capacitance of a hyperbole is approximated
in \cref{sec:hyperbole_capacitance}
by integrating the charge on the planar electrode,
\begin{equation}
    C_"H" (z)
        \propto
            \left({\ln \frac{4 z + 2 r_'p'}{r_'p'}} \right)^{-1}
            \,,
    \label{eq:C_hyperbole}
\end{equation}
where
$r_'p'$ is the tip curvature of the hyperboloid
and
$z$ is the distance to the plane.
The capacitance of a parallel plane capacitor is
\begin{equation}
    C_"P" (z)
        \propto \frac{1}{z}
        \,,
    \label{eq:C_plane}
\end{equation}
where
$z$ is the distance between the planes,
and the capacitance for a sphere above a plane is \citep{Crowley2008}
\begin{equation}
    C_"S" (z)
        \propto r_'p' \left( 1 + \frac{1}{2} \ln \left( 1 + \frac{r_'p'}{z} \right) \right)
        \,,
    \label{eq:C_sphere}
\end{equation}
where
$r_'p'$ is the radius of the sphere.
The difference in capacitance for the three models is substantial,
see \cref{fig:capacitance_models}.
A single sphere does not take a conducting channel into account,
and this is the reason why
its capacitance does not change significantly
before $z$ is about ten times $r_'p'$.
Conversely,
for the planar model, the
capacitance grows rapidly
as it doubles every time $z$ is halved,
but
assuming parallel planes is considered an extreme case.


\begin{figure}[t]
    \centering
    \includegraphics[width=0.95\linewidth]{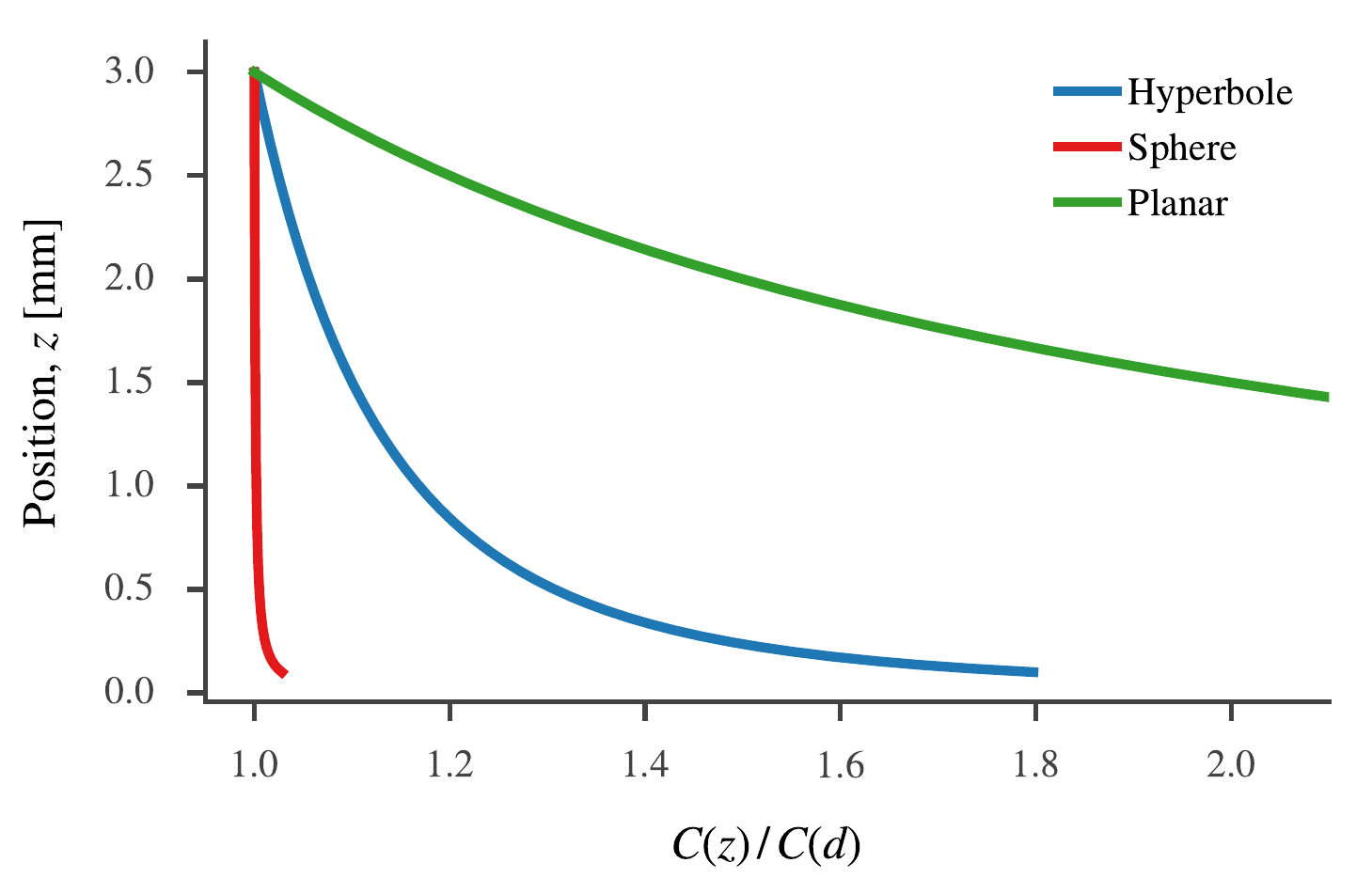}
    \caption{%
        The three proposed models for capacitance
        as a function of the position in gap.
        }
    \label{fig:capacitance_models}
\end{figure}

To test the impact of the variation in
streamer channel conductivity and capacitance on the streamer propagation
we will use a simplified model,
where
electrical breakdown within the channel is also included.
Each streamer head is assigned a
time constant $\tau$,
which is split into several contributions,
\begin{equation}
    \tau = fgh \tau_0
        \quad \text{with} \quad
    \tau_0 = \frac{C d}{A \sigma}
    \,,
    \label{eq:tau_model}
\end{equation}
where $d$ is the gap distance.
The contributions
\begin{equation}
    f = \frac{\ell}{d} \,,
    \quad
    g = \frac{C(z)}{C(d)} \,,
    \quad \text{and} \quad
    h = \Theta (E_'bd' - E_'s') \,,
    \label{eq:tau_model_factors}
\end{equation}
represent change in
resistance in the channel
\hl{($f$)},
capacitance
between
the streamer head
and the plane
\hl{($g$)},
and
the breakdown in the channel
\hl{($h$)},
respectively.
The Heaviside step function $\Theta$ is zero when
the electric field in the channel is larger than the breakdown threshold
($E_'s' > E_'bd'$)
and one otherwise.
When a breakdown in the channel occurs
$\Theta = 0$, giving $\tau = 0$,
and thus the potential at the streamer head is instantly
relaxed to the potential of the needle.
We therefore assume that breakdowns in the channel is the cause of re-illuminations.
Since the heads are individually connected to the needle,
a breakdown only affects one channel.

\hl{Having $\tau$ longer or shorter than the streamer propagation time
implies relatively low or high conductivity, respectively.
Since the contributions $f$, $g$, and $h$ are on the order of magnitude 1
for most parts of the gap,
the same is true for $\tau_0$
(although $\tau_0 = R(0) C(d)$ does not have a physical interpretation).}
Throughout the simulations,
an increase in $\tau_0$ is considered to arise from
a decrease in the channel conductivity $\sigma$, and vice-versa.
\hl{The influence of channel expansion (increasing $A$),
is included in the discussion in \mbox{\cref{sec:discussion}},
as well as evaluation of conductivity from $\tau_0$.}

\begin{figure*}
    \centering
    \includegraphics[width=0.95\linewidth]{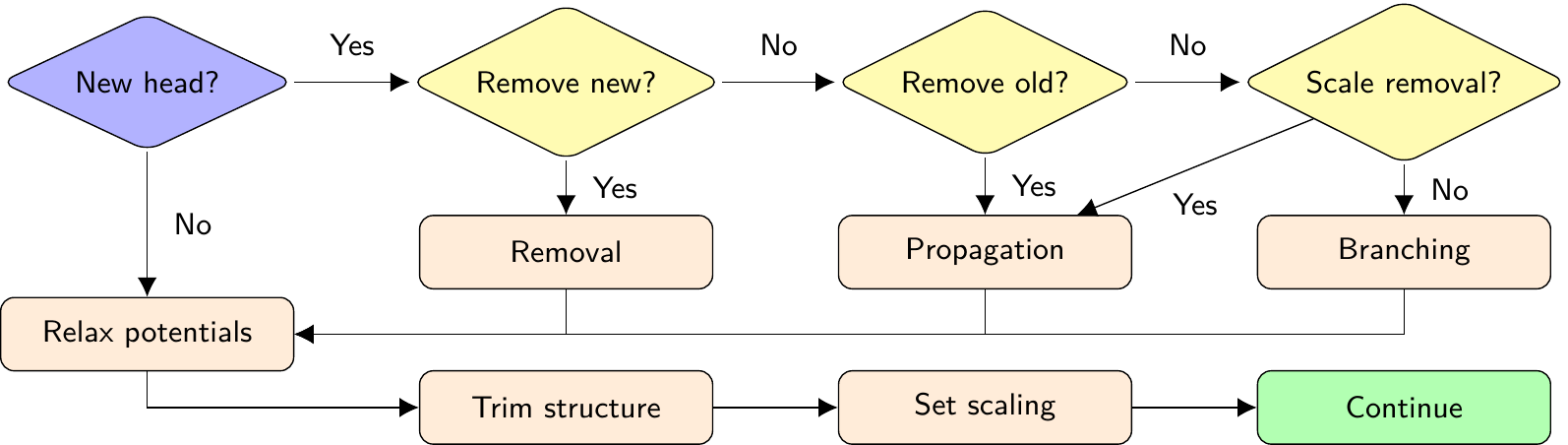}
    \caption{%
        Algorithm for updating the streamer structure.
        ``Collision'' and ``merging'' checks decides
        whether the head should be removed immediately.
        The same checks are then performed to see if the addition
        of the head causes an existing head to be removed.
        Then, a ``scale removal'' check is performed at equipotential.
        See text in \cref{sec:alg_streamer} for details on these checks.
        If the new head is not removed and
        a check finds a head to remove,
        the new head is a ``propagating'' head
        and its potential is set by~\cref{eq:charge_propagation}.
        Else,
        it is a ``branching head''
        and its potential is set by~\cref{eq:share_charge},
        which changes the potential of an existing streamer head as well.
        All potentials are then relaxed according to~\cref{eq:V_relax}.
        Finally, the structure is trimmed by checking
        ``collision'', ``merging'' and ``scale'',
        and the correct scale is set,
        as described in~\citep{Madshaven2018cxjf}.
        }
    \label{fig:alg_streamer}
\end{figure*}

\subsection{Electrical potential of new streamer heads}\label{sec:el_pot_new_head}

The potential of a new head $m$ is dependent on
the closest streamer head $n$ only.
This is an approximation
compared with using an electric network model for the streamer%
~\cite{Fofana1998bm4sd5}
and
in contrast to our previous model using fixed electric field in the streamer channel%
~\cite{Madshaven2018cxjf}.
Two different cases are implemented,
depending on whether the new head can cause a branching event or not
(see details in \cref{sec:alg_streamer} and \cref{fig:alg_streamer}).
If the new head is not considered to be a new branch
its potential is calculated assuming charge transfer from $n$ to $m$,
\begin{equation}
    V_m = V_n \frac{C_n}{C_m}  \,.
    \label{eq:charge_propagation}
\end{equation}
Secondly,
the potential for a branching head is calculated
by sharing the charge between $n$ and $m$,
reducing the potential of $n$ as well.
Isolating the two heads from the rest of the system,
the total charge is $Q = V_n C_n$,
and this charge
should be divided in such a way that
the heads obtain the same potential,
$V (\vec{r}_m) = V (\vec{r}_n)$,
using~\cref{eq:Vr_Er} for both $m$ and $n$.
Introducing $M_{ij} = V_j(\vec{r}_i) / V_j(\vec{r}_j)$,
\cref{eq:Vr_Er} is simplified as
\begin{equation}
    V_i(\vec{r}_i) = \sum\limits_j M_{ij} k_j V_j(\vec{r}_j)
    \;\Rightarrow\;
    1 = \sum\limits_j M_{ij} k_j
    \,,
\end{equation}
when all $V_i(\vec{r}_i)$ are equal.
The coefficients $k_j$ are obtained by \textsc{NNLS}-optimization~\cite{Madshaven2018cxjf},
like the potential shielding coefficients.
Finally,
the potential for both $m$ and $n$ is calculated as
\begin{equation}
    V_m (\vec{r}_m)
        = \frac{Q}{\sum k_i C_i}
        \,.
        \label{eq:share_charge}
\end{equation}
%
In the case where one $k_i$ is close to unity and the other is close to zero,
the result resembles~\cref{eq:charge_propagation},
however,
$\sum k_i \geq 1$,
so the potential will drop when the capacitance
of the new head is similar to or larger than its neighbor.
The potential of a new head
could also have been set to the potential at its position
calculated before it is added,
but that probably overestimates the reduction in potential,
since the avalanche itself distorts the electric field
and
since transfer of charge from neighboring heads
is faster than from the needle.

\subsection{Updating the streamer}\label{sec:alg_streamer}

In~\citep{Madshaven2018cxjf},
critical avalanches are replaced by new streamer heads and added to the streamer.
Any head within another head has ``collided'' with the streamer and is removed.
If two heads are too close to each other they are ``merged'',
implying that the one closest to the plane is kept and the other one is removed.
Also,
the potential shielding coefficients are calculated
and any head with a low coefficient is removed, ``scale removal''.
Finally,
the shielding coefficients are set.

The algorithm is now changed,
see \cref{fig:alg_streamer}
(replacing the block labeled ``Streamer''
in figure~5 in~\citep{Madshaven2018cxjf}).
New heads are either removed,
or classified as ``propagating'' or ``branching'',
and their potential is set using~%
\cref{eq:charge_propagation} or~\cref{eq:share_charge}.
If a head can be added without causing another to be removed,
it can cause a branching event,
else it represents propagation of the streamer.
The addition of one extra head is by itself not sufficient for streamer branching,
often there are several heads within one propagating branch.
Branching occurs through a process of
adding new heads to opposing sides of a cluster of heads
while removing the heads in the center
(cf. figure 28 in~{\cite{Madshaven2018cxjf}}).
With this approach,
branching follows as a consequence of propagation,
contrary to models in which streamers propagate by adding branches%
~\cite{Niemeyer1984d35qr4,Fofana1998bm4sd5}
or models which rely on inhomogeneities~{\cite{Jadidian2014f55gj5}}.

The difference in potential between each head $V_i$ and the needle $V_0$
is first found and then reduced,
\begin{equation}
    \Delta V_i = V_0 - V_i(\vec{r}_i)
    \; \rightarrow \;
    V_i(\vec{r}_i) = V_0 - \Delta V_i \, e^{ - \Delta t / \tau_i  }
    \,.
    \label{eq:V_relax}
\end{equation}
where the time constant of each head $\tau_i$ is calculated by~\cref{eq:tau_model}.
Finally,
the streamer structure is trimmed (collision, merge and scale removal)
and the potential scaling is optimized
as described in%
~\cite{Madshaven2018cxjf}.
Note trimming and rescaling is performed
to remove heads lagging behind
and
to ensure correct potential
at each streamer head,
however,
it does not preserve
charge and capacitance.
For this reason,
we do not calculate the total charge or capacitance of the streamer.

}


\section{Single channel streamer at constant speed}\label{sec:single}{

\begin{figure}[t]
    \centering
    \includegraphics[width=0.98\linewidth]{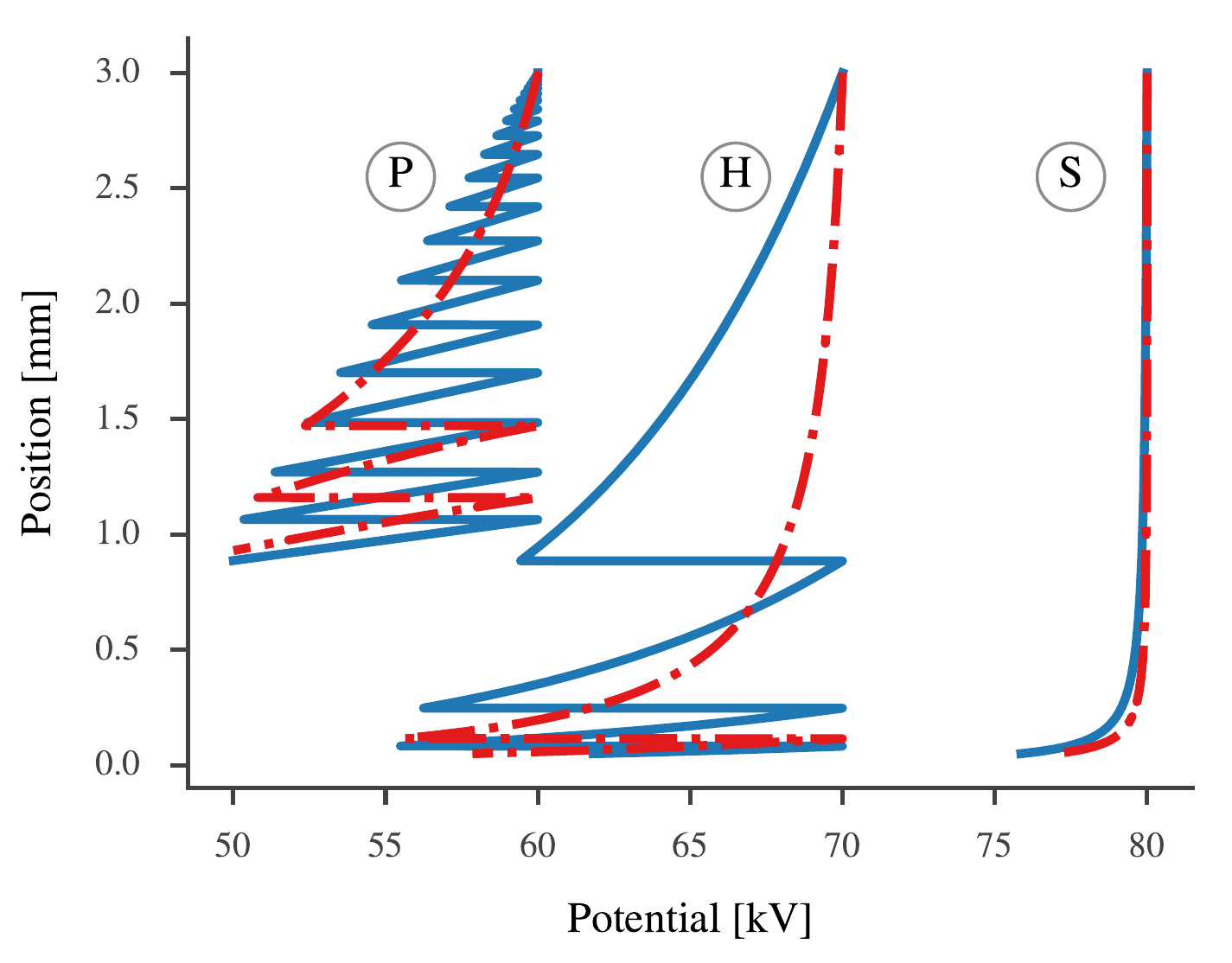}
    \caption{%
        Potential of single head propagating at constant speed,
        starting at at different potentials
        for different capacitance models:
        (P)lane, (H)yperbole, and (S)phere,
        Time constants
        $\tau_0 = \SI{0.1}{\micro \second}$ (dashed red)
        and
        $\tau_0 = \SI{10}{\micro \second}$ (solid blue),
        and breakdown in channel at
        $E_'bd' = \SI{5}{\kilo \volt \per \milli \metre}$.
        }
    \label{fig:simple_model}
\end{figure}

As a model system,
a simplified numerical model is investigated
by considering a streamer propagating
as a single branch at constant speed.
The parameters used are
gap distance $d = \SI{3}{\milli \metre}$,
propagation speed
$v_'p' = \SI{3}{\kilo \metre \per \second}$,
tip radius $r_'p' = \SI{6}{\micro \metre}$,
minimum propagation voltage $V_'p' = \SI{50}{\kilo \volt}$,
and
breakdown in the channel at $E_'bd' = \SI{5}{\kilo \volt \per \milli \metre}$.
The time constant $\tau$ is modeled by~\cref{eq:tau_model,eq:tau_model_factors}
and the potential is calculated by~\cref{eq:charge_propagation}.

The result of varying $\tau_0$
for the different capacitance models $g$,
is shown in \cref{fig:simple_model}.
When applying the sphere model in~\cref{eq:C_sphere},
the change in potential is small
and the time constant has little influence,
as expected based on \cref{fig:capacitance_models}.
The potential changes faster with the hyperbole model in~\cref{eq:C_hyperbole}
and breakdown in the channel occurs in the final part of the gap.
Decreasing $\tau_0$, i.e.\ increasing the conductivity,
reduces the potential drop
and delays the onset of breakdowns in the channel.
This is similar for the plane model in~\cref{eq:C_plane},
where rapid breakdowns at the start of the propagation
are suppressed by decreasing $\tau_0$.
The propagation for the plane model is stopped
when the potential drops below $V_'p'$,
which occurs at about the same position for both low and high $\tau_0$.
Where the propagation stops depends on the capacitance model,
the breakdown in channel threshold,
the time constant,
and
the initial voltage $V_0$.
A reduction of $V_0$ by $\SI{10}{\kilo \volt}$
for the hyperbole model
would have stopped these streamers as well,
but the one with higher conduction would have propagated most of the gap,
stopping close to the opposing electrode.

By assuming an initial capacitance
$C = {\SI{0.1}{\pico \farad}}$,
the energy ($W = \frac{1}{2} C V^2 = \frac{1}{2} Q^2 C^{-1}$)
of each streamer head in {\cref{fig:simple_model}}
is some hundred {$\si{\micro\joule}$}.
From {\cref{fig:capacitance_models}},
the capacitance of the hyperbole model
increases by about 20~\% during the first {$\SI{2}{\milli\meter}$},
which amounts to some tens of {$\si{\micro\joule}$},
and more than approximately {$\SI{5}{\micro\joule\per\milli\meter}$}
required for propagation~{\cite{Gournay1994dw59f5}}.
Just before the first breakdown
for the low-conductivity ``hyperbole streamer'' in {\cref{fig:simple_model}},
there is a voltage difference of about {$\SI{10}{\kilo\volt}$}.
Given a $\tau$ of about {$\SI{10}{\micro\second}$}
this equals a continuous current of about {$\SI{100}{\micro\ampere}$},
while the first breakdown adds about a {$\si{\nano\coulomb}$} of charge.
In comparison, the high-conductivity ``hyperbole streamer''
has a current of more than a {$\si{\milli\ampere}$},
sustaining the potential at the streamer head for the first part of the propagation.
As such, the current and charge are comparable to experimental results%
~{\cite{Ingebrigtsen2009fptpt5}}.

}


\section{Numerical simulation results}{\label{sec:results}

\begin{table}[!t]
    \caption{
        Model parameter values.
        }
{%
\centering
\renewcommand{\arraystretch}{1.1}%
\begin{tabularx}{1.0\linewidth}{
    >{\raggedright}X
    >{\centering\arraybackslash\hsize=.15\hsize}X
    >{\centering\arraybackslash\hsize=.40\hsize}X
    }%
    \toprule
        Gap distance
    &   $d_'g'$
    &   $\SI{3.0}{\milli\metre}$
    \\
        Needle curvature
    &   $r_'n'$
    &   $\SI{6.0}{\micro\metre}$
    \\
        Streamer head curvature
    &   $r_'s'$
    &   $\SI{6.0}{\micro\metre}$
    \\
        Scattering constant
    &   $E_\alpha$
    &   $\SI{1.9}{\giga\volt\per\metre}$
    \\
        Max avalanche growth
    &   $\alpha_'m'$
    &   $\SI{130}{\per\micro\metre}$
    \\
        Meek constant
    &   $Q_'c'$
    &   $\SI{23}{ }$
    \\
        Electron mobility
    &   $\mu_'e'$
    &   $\SI{45}{\milli\metre^{2}\per{\volt\second}}$
    \\
        Anion number density
    &   $n_'ion'$
    &   $\SI{2e12}{/\metre^{3}}$
    \\
        Head merge threshold
    &   $d_'m'$
    &   $\SI{50}{\micro\metre}$
    \\
        Shielding threshold
    &   $k_'c'$
    &   $\SI{0.10}{}$
    \\
        Simulation time step
    &   $\Delta t$
    &   $\SI{1.0}{\pico\second}$
    \\
    \bottomrule%
\end{tabularx}%
}%

    \label{tab:tab_param}
\end{table}

Positive streamers in cyclohexane are simulated
in a needle-plane gap.
Model parameters discussed in this work are given in~{\cref{tab:tab_param}}.
The base parameters and their influence on the model
were discussed in~{\cite{Madshaven2018cxjf}}
and is therefore not repeated here.
The values for $\alpha_'m'$ and $E_\alpha$
have been taken from~{\cite{Naidis2015gbf7x2}} rather than~{\cite{Haidara1991dk77wh}},
decreasing the propagation voltage
from about $\SI{60}{\kilo\volt}$ to about $\SI{40}{\kilo\volt}$~{\cite{Madshaven2018cxjf}},
which is closer the experimentally estimated $\SI{33}{\kilo\volt}$~{\cite{Ingebrigtsen2009fptpt5}}.
\hl{Experimentally,
the propagation voltage is determined from either
the streamer shape,
the measured current,
or interpolation of the propagation length
~\mbox{\cite{Gournay1993ck4hxg,Ingebrigtsen2009fptpt5}}.
For our simulations investigating propagation,
however,
the minimum requirement is simply
a streamer length of 25 \% of the gap,
since most simulated non-breakdown streamers stop
within the first few hundred {\si{\micro\meter}}%
~\mbox{\cite{Madshaven2018cxjf}}.}
In the updated model,
the field in the streamer
$E_'s'$ is not fixed but calculated
by applying the RC-model described here.
The influence of the
conduction and breakdown in the streamer channel
is investigated by changing values for
$\tau_0$ and $E_'bd'$.
\hl{Interesting values for $\tau_0$ are within some orders of magnitude
of the propagation time for a streamer.
The interpretation in terms of streamer radius and conductivity
is discussed in the next section.
For $E_'bd'$ to affect streamers in a {\si{mm}}-sized gap,
minimum some {\si{kV/mm}} are needed,
however,
the average electric field within the streamer $E_'s'$
is dependent on both $\tau_0$ and $E_'bd'$.}
%
In \cref{sec:single},
we indicate how conductivity and capacitance
influence the potential of a streamer propagating at constant speed.
In this section,
however,
only the hyperbole model for capacitance is used.
Furthermore,
the propagation speed depends on the potential in the simulation model%
~\cite{Madshaven2018cxjf},
and
allowing multiple heads increases the total capacitance of the streamer,
which gives a drop in potential when an extra streamer head is added.

The simulations presented in
\cref{fig:multi_trail_streak_epot}
have equal voltage
and equal initial anion placement (initial random number).
The streamers are visualized in \cref{fig:multi_trail_streak_epot}(a),
showing some increase in thickness and decrease in branching
when the conductivity increases,
however,
their propagation speeds in \cref{fig:multi_trail_streak_epot}(b) clearly differ.
The propagation speed is mainly influenced by
the number of streamer heads
and
the potential of the streamer heads%
~\citep{Madshaven2018cxjf}.
\Cref{fig:multi_trail_streak_epot}(c)
shows that when
there is no breakdown in the channel,
and
the conductivity is low,
i.e.\ $\tau_0$ is high compared to the gap distance and propagation speed,
the potential is reduced as the streamer propagates.
For some short distances,
the slow potential reduction is similar to the results in \cref{fig:simple_model},
however,
when an extra head is added to the streamer (possible branching)
there is a distinct reduction in the potential of some $\si{\kilo \volt}$.
Increased conductivity
increases the speed and average potential of the streamers
in \cref{fig:multi_trail_streak_epot}(c).
At
$\tau_0 = \SI{e-6}{\second}$,
a single branch may gain potential during propagation
while branching reduces the overall potential.
This is reasonable since $\tau_0$ is about a tenth of the time to cross the gap,
see \cref{fig:multi_trail_streak_epot}(b).
By further increasing the conductivity
(decreasing $\tau_0$ to $\SI{e-8}{\second}$),
the potential is kept close to that of the needle
and
the speed is increased,
but $\tau_0$ is now less than a hundredth of the time to cross,
implying that it has little influence on the simulation.

\begin{figure*}[t]
    \centering
    \includegraphics[width=0.49\textwidth]{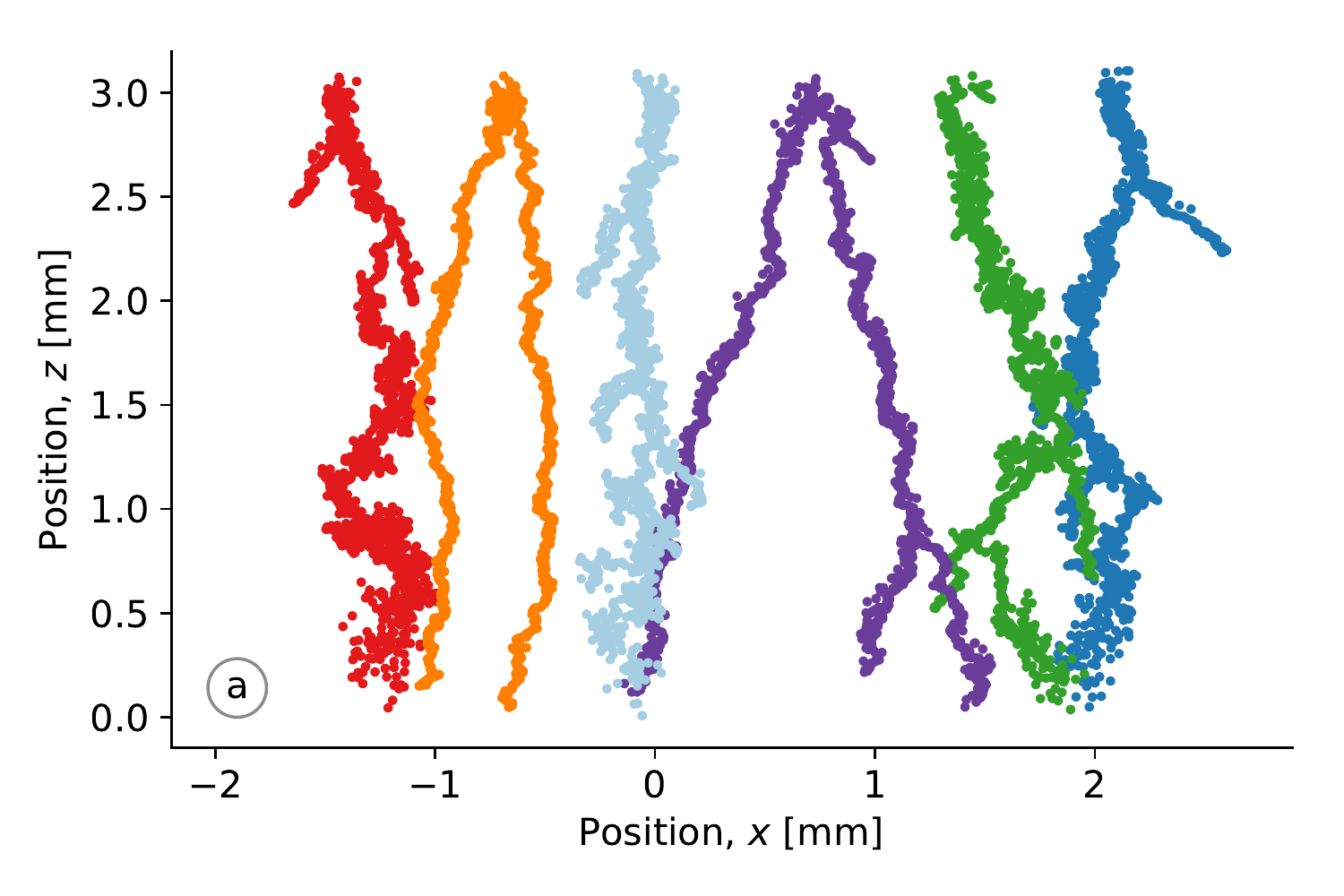}
    \includegraphics[width=0.49\textwidth]{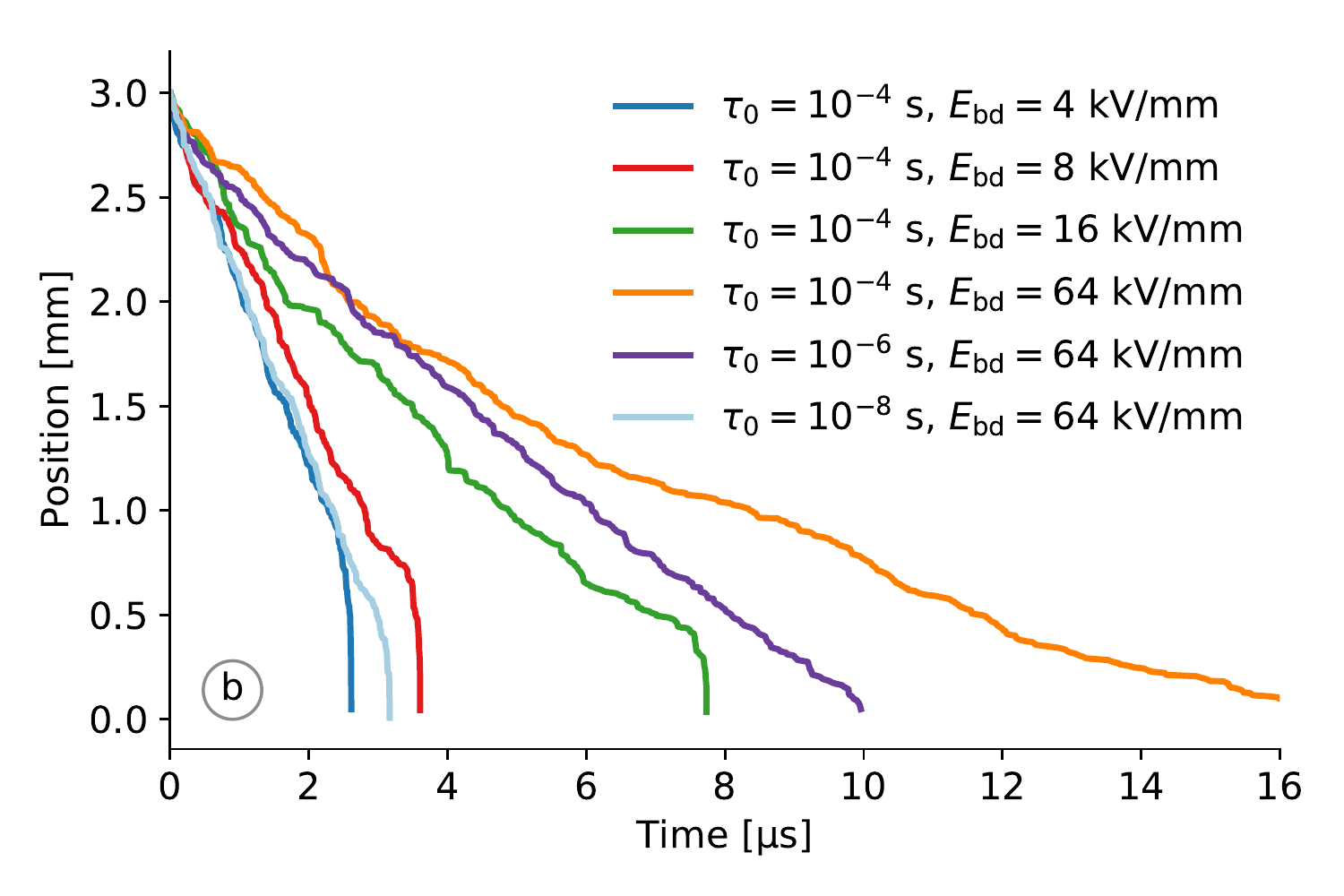}
    \includegraphics[width=0.49\textwidth]{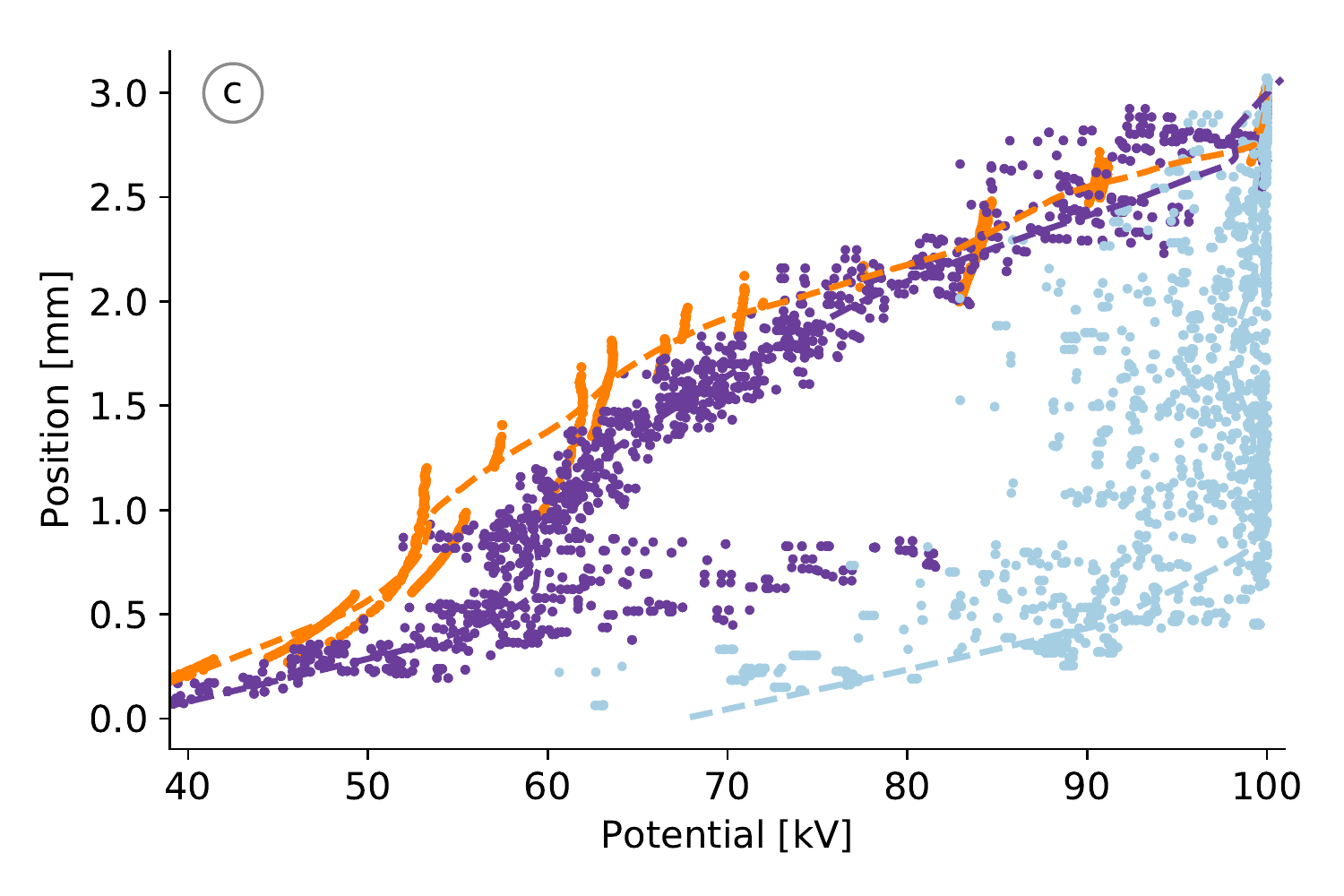}
    \includegraphics[width=0.49\textwidth]{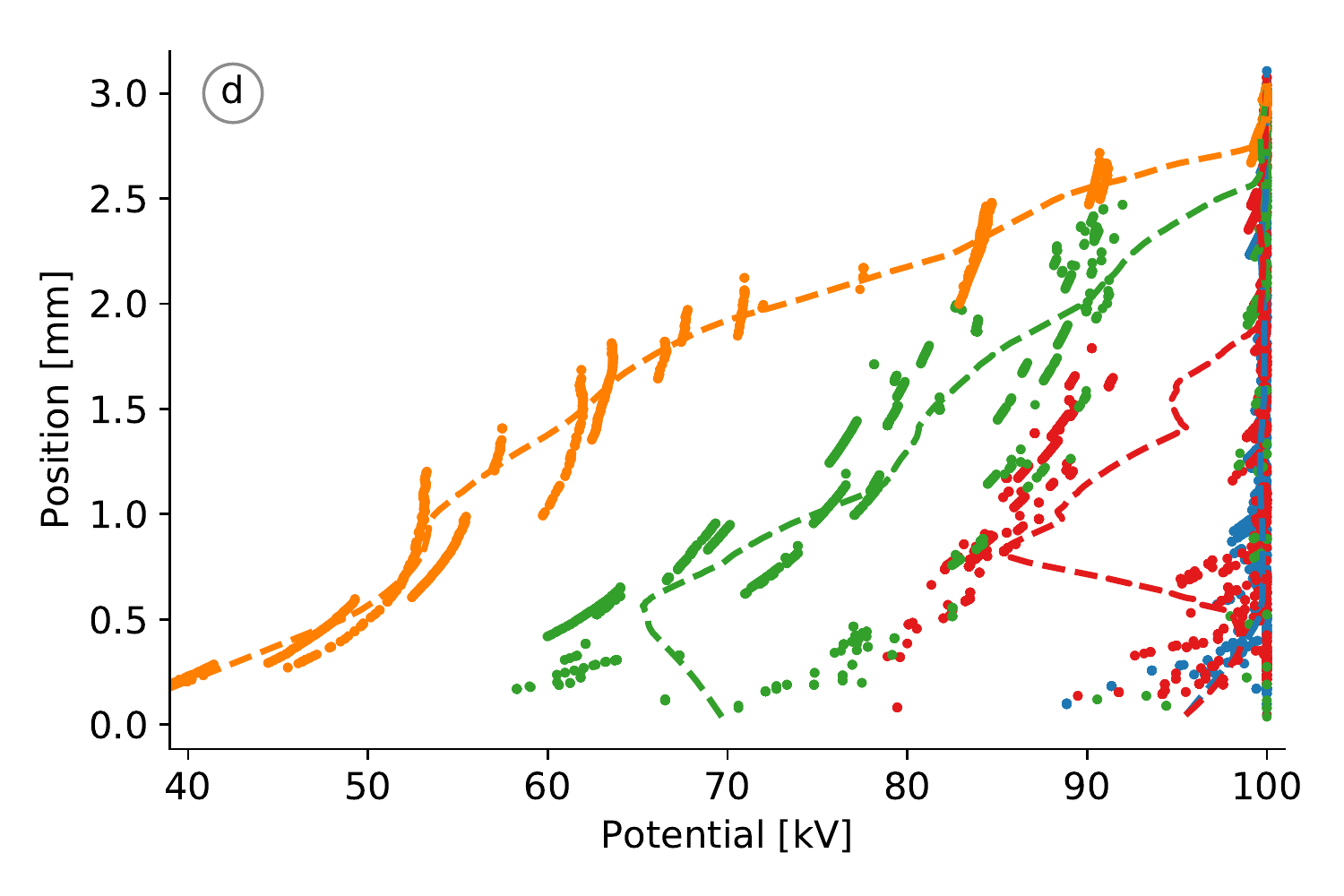}
    \caption{%
        Simulations carried out at $\SI{100}{\kilo \volt}$
        using the same initial anion placement
        for a number of
        time constants $\tau_0$
        and
        breakdown thresholds $E_'bd'$.
        (a) ``Shadowgraphic plot'' where the position of each streamer head is marked
        and
        (b) ``streak plot'' showing the leading streamer head vs time.
        (c) and (d) show the potential of streamer heads vs position,
        \hl{and can be compared with \mbox{\cref{fig:simple_model}}.
        They show how decreasing
        $\tau_0$ or $E_'bd'$, respectively, increases the average potential.
        Dots close to $\SI{100}{\kilo \volt}$ in (d) indicate a recent channel breakdown (re-illumination).}
        In (d), $E_'bd' = \SI{4}{\kilo \volt \per \milli \metre}$
        is close to maximum and mostly hidden behind the others.
        The dashed lines are moving averages.
        \hl{%
        All streamer heads involved in each simulation is shown in (a),
        only the leading head is shown in (b).
        In (c) and (d),
        data is sampled every {\SI{3}{\micro \meter}} of the propagation.
        Each dot in (b), (c) and (d) is also shown in (a), but not vice-versa.}
        }
    \label{fig:multi_trail_streak_epot}
%
\vspace*{2 ex}
%
    \centering
    \includegraphics[width=0.49\textwidth]{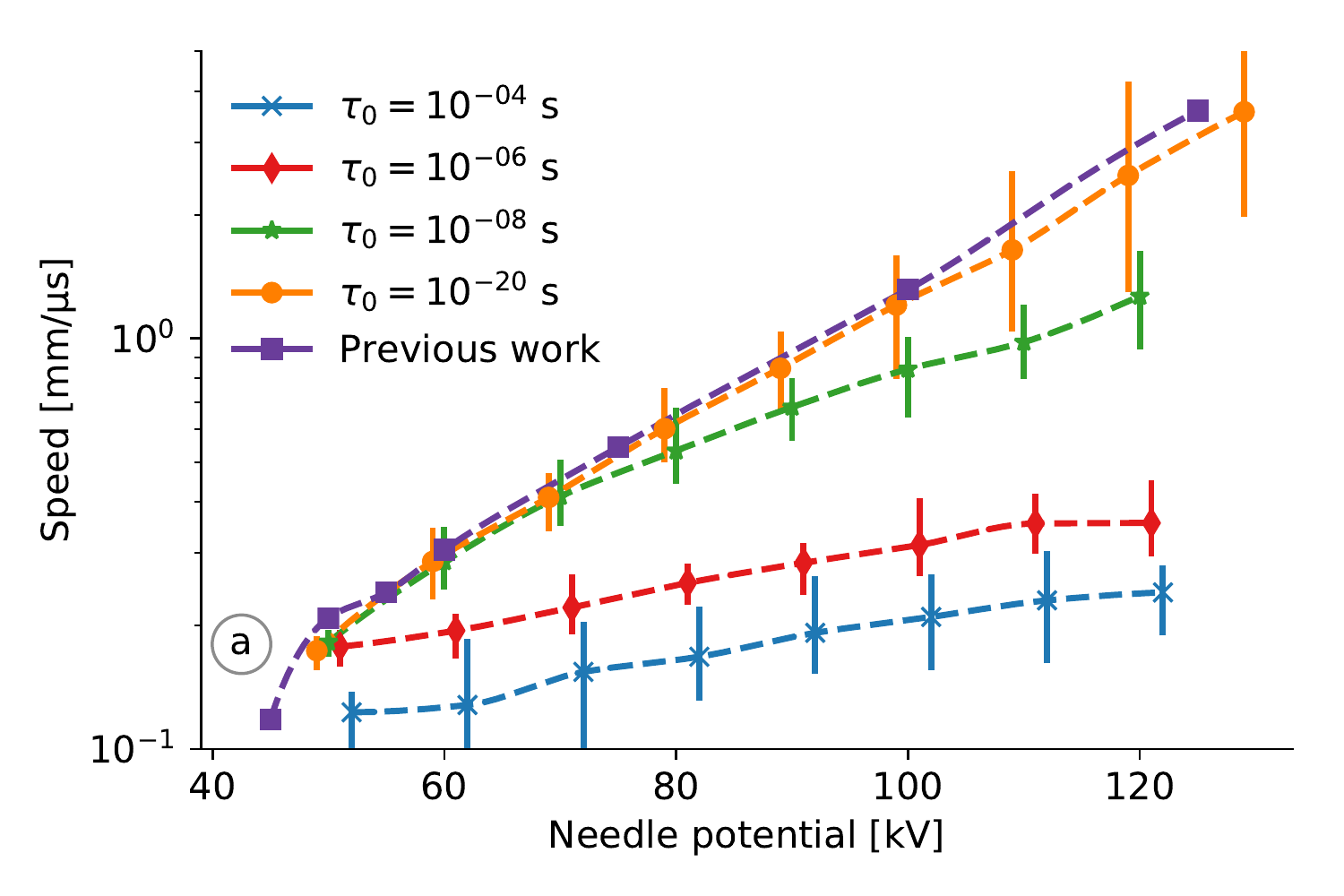}
    \includegraphics[width=0.49\textwidth]{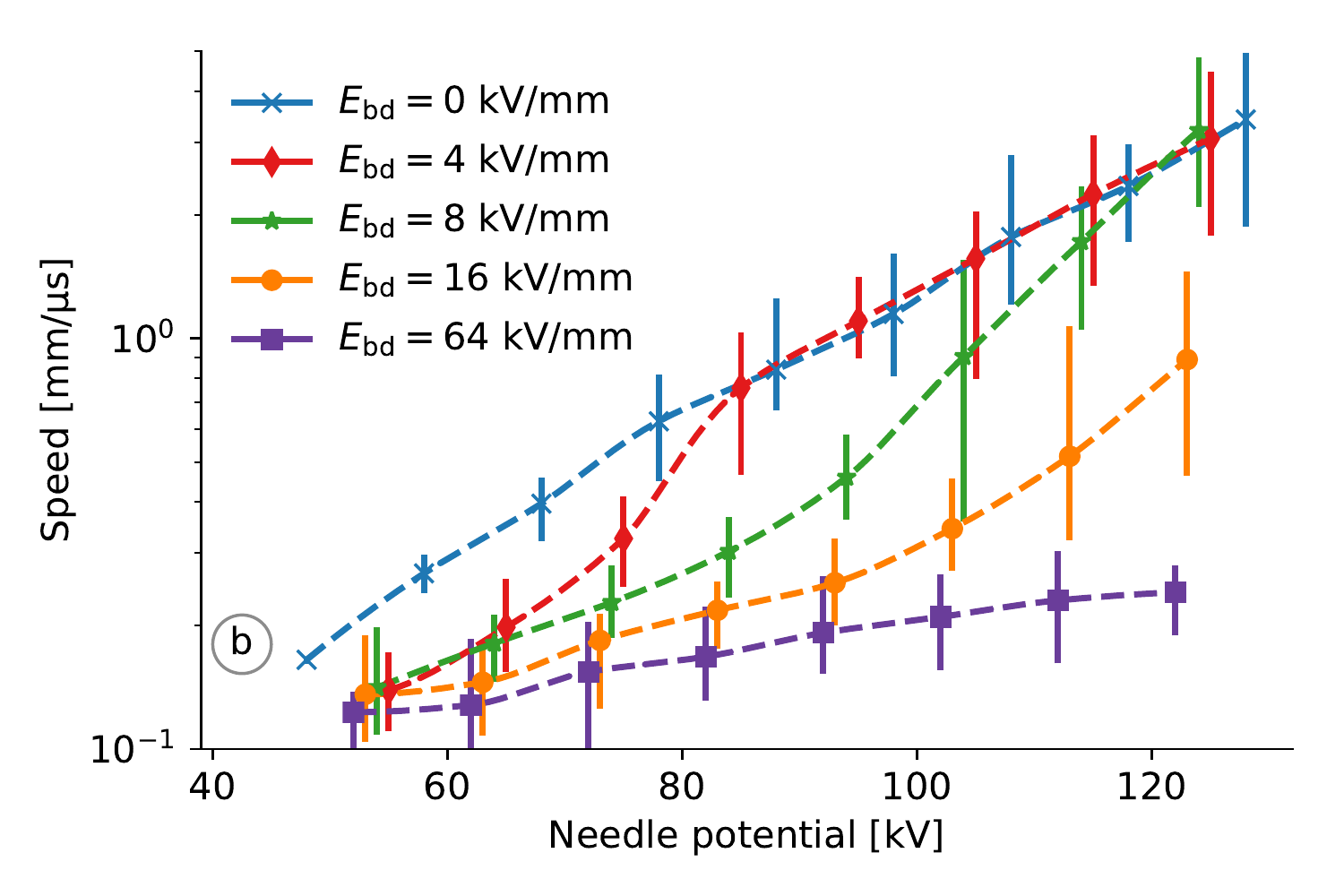}
    \caption{%
        Propagation speed calculated for the mid
        $\SI{1.5}{\milli \metre}$ of the gap.
        (a) for different time constants $\tau_0$
        with $E_'bd' = \SI{64}{\kilo \volt \per \milli \metre}$,
        and
        (b) for different breakdown thresholds $E_'bd'$
        with $\tau_0 = \SI{e-4}{\second}$.
        Twenty simulations
        are performed for each voltage,
        the dashed lines are interpolated to the average values
        and
        the bars cover the minimum and maximum values.
        ``Previous work'' is data from~\cite{Madshaven2018cxjf}
        (figure~15, $E_\alpha = \SI{2}{\giga\volt \per \metre}$).
        Simulations where
        $\tau_0 = \SI{e-20}{\second}$
        or
        $E_'bd' = \SI{0}{\kilo \volt \per \milli \metre}$
        are comparable to our previous work since
        $\tau$ is effectively zero for all of them.
        \hl{Each simulation is initiated
        with a random number
        to ensure that the configurations of seeds are uncorrelated.}
        }
    \label{fig:zapsim_stat_v_psm_rct-rcb}
\end{figure*}

For low channel conductivity,
there is less ``scatter'' in the streamer potential,
which makes it easier to interpret the results
when investigating the effect of breakdown in the streamer channel,
see
\cref{fig:multi_trail_streak_epot}(d).
Breakdown in the channel can occur in the first part of the gap
even when the threshold $E_'bd'$ is high,
since a potential difference of some $\si{\kilo \volt}$
gives an electric field of several $\si{\kilo \volt \per \milli \metre}$
when the streamer length is some hundred $\si{\micro \metre}$.
For $E_'bd' = \SI{16}{\kilo \volt \per \milli \metre}$
in \cref{fig:multi_trail_streak_epot}(d),
the average field inside the streamer is about
$\SI{13}{\kilo \volt \per \milli \metre}$. 
Rapid breakdowns gives $E_'s'$ close to zero
for $E_'bd' = \SI{8}{\kilo \volt \per \milli \metre}$,
except for about $\SI{0.5}{\milli \metre}$
in the middle of the gap.
The average field in a streamer is
on the order of $\si{\kilo \volt \per \milli \meter}$%
~\cite{Saker1996}.
It is seen in
\cref{fig:multi_trail_streak_epot}(b)
that the streamer slows down
for the portion of the gap
where the potential is decreased,
and that streamers having similar average potential
also use similar times to cross the gap.

\hl{\mbox{\Cref{fig:multi_trail_streak_epot}}
gives a good qualitative indication of how
$\tau_0$ and $E_'bd'$ affects the simulations.
Different initial configuration
of seed electrons show similar trends.
Increasing concentration of seeds increases streamer propagation speed,
but not branching~{\cite{Madshaven2018cxjf}}.
However,
changing}
the initial configuration changes the entire streamer breakdown
and adds stochasticity to the model,
while changing the needle voltage
influences most results,
such as
the propagation speed,
the jump distances,
the number of branches,
and
the propagation length%
~\cite{Madshaven2018cxjf}.
The effect of $\tau_0$ and $E_'bd'$ on the propagation speed
is shown in \cref{fig:zapsim_stat_v_psm_rct-rcb}
for a range of voltages,
with several simulations
performed at each voltage.
The simulations with the lowest $\tau_0$
are similar to those with the lowest $E_'bd'$.
For these simulations,
the potential of the streamer is equal to the potential of the needle,
and
the results are similar to those
presented in figure~15 in~\citep{Madshaven2018cxjf},
as expected.
Increasing $\tau_0$ can reduce the propagation speed
for a given voltage,
and the time constant seems to dampen the
increase in speed following increased voltage.
Adding the possibility of a breakdown in the channel
reverses this,
since the net effect is a reduction in the average time constant,
i.e.~an increase in net conductivity.
At low needle potential,
there are fewer breakdowns in the channel
and the speed is mainly controlled by the conductivity through $\tau_0$,
however,
breakdowns become more frequent with increasing needle potential,
which in turn increase the streamer potential and speed.

}


\section{Discussion}\label{sec:discussion}{

As for our original model~\cite{Madshaven2018cxjf},
this updated model still predicts
a low propagation speed
(see {\cref{fig:zapsim_stat_v_psm_rct-rcb}})
and a low degree of branching
(see {\cref{fig:multi_trail_streak_epot}(a)})
compared with experimental results%
~{\cite{Ingebrigtsen2007ck6k6q,Ingebrigtsen2009fptpt5}}.
Low propagation speed can be caused by
low electron mobility,
low electron/anion seed density,
or too high shielding between streamer heads%
~{\cite{Madshaven2018cxjf}}.
Increasing the time constant seems to increase the number of branches
by regulating their speed
and
introducing breakdown in the channel reverses this effect.
The hyperbole approximation of the electric field
gives a strong electric field directed towards the planar electrode.
Thus, electron avalanches in front of the head,
giving forward propagation is favored over off-axis propagation
and the chance of branching is reduced.
A hyperbole can be a good approximation
in the proximity of a streamer head,
while possibly overestimating the potential in regions farther away.
An overestimation of the potential from the streamer heads
results in lower $k_i$ values for the heads,
which in turn gives
lower electric fields,
slower streamers,
and
a higher probability of a branch stopping,
especially for branches lagging behind the leading head.
%
Since we model an ``infinite'' planar electrode,
the capacitance does not change with the $xy$-position of an individual branch
(unlike e.g.~{\cite{Fofana1998bm4sd5}}).
The coefficients $k_i$ scale the streamer heads
when the electric potential from the streamer is calculated,
and changing a $k_i$ can be interpreted as changing the capacitance
of a streamer head.
Two heads give a streamer a higher capacitance,
but not twice the amount of a single head.
However,
the scaling is calculated from the potential and not the geometry,
so this interpretation is an approximation,
and for this reason we do not explicitly calculate the
total capacitance or injected charge from the electrodes.
The total injected current will reflect
the behavior of individual heads
discussed in {\cref{sec:single}},
having both a continuous component and impulses following breakdowns.

The conductivity of the channels can be approximated from the time constants.
Consider that
$d = \SI{3}{\milli \metre}$,
$C = \SI{0.1}{\pico \farad}$,
and
$A = \SI{100}{\micro \metre^2}$,
results in that
$\sigma = \SI{3}{\siemens \per \metre}$
is required for
$\tau_0 = \SI{1}{\micro \second}$
according to~\cref{eq:tau_model}.
{\Cref{fig:zapsim_stat_v_psm_rct-rcb}} thus shows that
a conductivity of some $\si{\siemens\per\metre}$
regulates the propagation speed,
and that increased conductivity increases the speed.
This is the order of magnitude
as estimated for the streamer channel \cite{Torshin1995}
and used by other models%
~{\cite{Fofana1998bm4sd5,Aka-Ngnui2006b8dr5t}},
which is a very high conductivity compared with the liquid
(about $\SI{e-13}{\siemens \per \meter}$~\cite{Lesaint2016cxmf}).
A streamer propagating at $\SI{1}{\kilo \metre \per \second}$
bridges a gap of $\SI{1}{\milli \metre}$
in $\SI{1}{\micro \second}$,
which implies that $\tau$ has to be shorter than this
to have a significant effect on the propagation,
in line with the results in \cref{fig:zapsim_stat_v_psm_rct-rcb}.
However, how frequent and how large the loss in potential is
as the streamer propagates, is also important in this context.

The streamer model permits a streamer branch to propagate
with a low reduction in potential,
enabling a branch to propagate a short distance
even when the channel is non-conducting.
However,
propagation
and
branching events increases the capacitance,
which reduces the potential at the streamer head,
and can result in a breakdown in the channel,
i.e.\ a re-illumination.
A re-illumination increases the potential of the streamer head,
possibly causing other branches to be removed,
and increases the chance of a new branching.
A breakdown in the channel of one streamer head
does not cause the nearby heads to increase in potential
since each streamer head is individually ``connected'' to the needle
(see \cref{fig:streamer_rc}).
\hl{Streamer experiments sometimes show re-illumination of single branches%
~{\cite{Linhjell1994chdqcz}},
but often more than one branch light up at the same time,
which is a limitation in the present model.}
Such effects can be investigated by
further development towards an electric network model
for the streamer channels and streamer heads~\cite{Fofana1998bm4sd5}.

A streamer channel is not constant in size,
but grows and collapses dynamically~\cite{Kattan1989}.
This implies that $A$ in~\cref{eq:R} changes with time,
but so does $\sigma$,
which depends on the density and mobility of the charge carriers.
In turn,
the creation, elimination, and mobility of the charge carriers
is dependent on the pressure in the channel.
\hl{%
Hence,
it is not straightforward
to evaluate how the conductivity of the channel is affected by the expansion.
Conversely,
external pressure reduces the diameter of the streamer channels%
~{\cite{Gournay1994dw59f5}},
and reduce stopping lengths
without affecting the propagation speed~{\cite{Lesaint1994d5q8x4}}.
In a network model,
each zigzag in each branch
can be assigned specific parameters
allowing greater control of the individual parts of the streamer,
such as channel radius and conductivity.
In the current implementation of the model,
the channel length calculation
and the constant conductivity
(except for breakdowns),
are aspects that can be improved in the future.
Accounting for the actual length of the streamer channel
is a minor correction,
whereas branched streamer heads ``sharing'' parts of a channel
can influence the simulation to a larger degree.}

From
\cref{sec:single}
we find that a channel with high conductivity
has less frequent re-illuminations,
in line with experiments \cite{Dung2012czgj}.
The results in
\cref{fig:simple_model}
and \cref{fig:multi_trail_streak_epot}
also indicate
that even with a collapsed channel
(where low/none conductivity is assumed)
a streamer is able to propagate some distance.
Whereas experiments indicate that,
1st mode streamers may propagate only a short distance
after the channel disconnects from the needle%
~\cite{Costeanu2002fhc926},
but the stopping of second mode streamers
occur prior to the channel collapsing%
~\cite{Gournay1994dw59f5}.
In our model,
restricting the conductivity
reduces potential in the extremities of the streamer
as the streamer propagates,
which regulates the propagation speed
and increases branching (\cref{fig:zapsim_stat_v_psm_rct-rcb}).
The potential is reduced until either
the streamer stops,
the propagation potential loss is balanced by conduction,
or a re-illumination occurs
and temporary increases the conductivity.
This seems to contrast experimental results where
the propagation speed of 2nd mode streamers is
just weakly dependent on the needle potential~\cite{Lesaint1994d5q8x4}
and re-illuminations does not change the speed~\cite{Dung2012czgj}.
However,
whether a channel is ``dark'' or ``bright''
can affect the propagation speed of higher modes~\cite{Lu2016f866wd}.

}


\section{Conclusion}\label{sec:conclusion}{

We have presented an RC-model
which includes conductivity and capacitance of the streamer.
This model has been applied in combination
with a streamer propagation model
based on the avalanche mechanism~%
\cite{Madshaven2018cxjf}.
The RC-model introduces a time constant
that regulates the speed of streamer propagation,
depending on the conductivity of the channel
and
the capacitance in front of of the streamer.
The streamer can propagate even when the channels are
non-conducting,
but then with reduction in potential
which reduces the speed and may cause stopping.
However,
re-illuminations, breakdowns in the channel, increase its conductivity
and the speed of the streamer.
It is also found that streamer branching,
which increases the capacitance and reduces the potential
at the streamer heads,
can give rise to re-illuminations.
Some limitations of our previous model~\cite{Madshaven2018cxjf},
such as the low propagation speed
and low degree of branching,
are not significantly affected by the addition of the RC-model,
and need to be investigated further.

}


\section*{Acknowledgment}{
The work has been supported by
The Research Council of Norway (RCN),
ABB and Statnett,
under the RCN contract 228850.
The authors would like to thank
Dag Linhjell
and
Lars Lundgaard
for
interesting discussions
and
for sharing their knowledge on streamer experiments.

}

\appendix

\section{Hyperbole capacitance}\label{sec:hyperbole_capacitance}{

The electric field from a hyperbole is~\cite{Madshaven2018cxjf}
\begin{equation}
    E = \frac{c}{a \sin \nu \sqrt{\sinh^2 \mu + \sin^2 \nu}} \,,
\end{equation}
where
$c$ and $a$ are constants given by the potential and the geometry,
and $\mu$ and $\nu$ are prolate spheroid coordinates.
In the $xy$-plane,
$\sin \nu = 1$
giving
$\sinh^2 \mu +1 = \cosh^2 \mu$,
and $E$ becomes a function of the radius $r$,
\begin{equation}
    E
        = \frac{c}{a \cosh \mu}
        = \frac{c}{\sqrt{r^2 + a^2}}
    \,,
\end{equation}
by using relations from~\cite{Madshaven2018cxjf}.
The charge $Q$ of a system is given by
the capacitance $C$
and
the potential $V$
through
$Q = CV$.
The charge of the hyperbole
is equal to
the charge on the surface electrode,
which
is found by integration of the electric field
using Gauss' law
\begin{equation}
    Q = 2 \pi \epsilon_0 \int\limits_0^R E \, r \,\d r
         = 2 \pi \epsilon_0 c \left( \sqrt{R^2 + a^2} -a \right)
         \,,
\end{equation}
where
$\epsilon_0$ is the vacuum permittivity.
Implying that $Q \propto c$
for a plane of a finite radius $R \gg a$.
From \citep{Coelho1971},
$\left. c \approx {2 V} \smash{\big/} {\ln (4 a / r_'p')} \right.$
and by using $a = z + \frac{1}{2} r_'p'$,
we find an expression for the capacitance of a hyperbole
\begin{equation}
    C_"h" = \frac{Q}{V}
        \propto \frac{c}{V}
        = 2 \left( {\ln \frac{4z + 2 r_'p'}{r_'p'}} \right)^{-1}
        \,,
\end{equation}
which depends on
the tip curvature $r_'p'$
and
the distance from the plane $z$.

}


\end{document}